\documentclass[conference]{IEEEtran}
\IEEEoverridecommandlockouts
\usepackage{cite}
\usepackage{amsmath,amssymb,amsfonts}
\usepackage{algorithmic}
\usepackage{graphicx}
\usepackage{textcomp}
\usepackage{xcolor}
\def\BibTeX{{\rm B\kern-.05em{\sc i\kern-.025em b}\kern-.08em
    T\kern-.1667em\lower.7ex\hbox{E}\kern-.125emX}}
\begin{document}

\newcommand{\shrink}[1][1]{\vspace{-#1\dimexpr0.25cm\relax}}
\newcommand{\shrinklittle}{\vspace{-0.10cm}}

\title{A Data-driven Analysis of Code Optimizations}

\author{\IEEEauthorblockN{Yacine Hakimi}
\IEEEauthorblockA{\textit{Ecole Superieure d'Informatique} \\
Algiers, Algeria \\
y\_hakimi@esi.dz}
\and
\IEEEauthorblockN{Riyadh Baghdadi}
\IEEEauthorblockA{\textit{New York University Abu Dhabi} \\
Abu Dhabi, UAE \\
Baghdadi@nyu.edu}
}

\maketitle

\begin{abstract}
As the demand for computational power grows, optimizing code through compilers becomes increasingly crucial. In this context, we focus on fully automatic code optimization techniques that automate the process of selecting and applying code transformations for better performance without manual intervention. Understanding how these transformations behave and interact is key to designing more effective optimization strategies. Compiler developers must make numerous design choices when constructing these heuristics. For instance, they may decide whether to allow transformations to be explored in any arbitrary order or to enforce a fixed sequence. While the former may theoretically offer the best performance gains, it significantly increases the search space. This raises an important question: Can a predefined, fixed order of applying transformations speed up the search without severely compromising optimization potential? In this paper, we address this and other related questions that arise in the design of automatic code optimization algorithms. Using a data-driven approach, we generate a large dataset of random programs, apply random optimization sequences, and record their execution times. Through statistical analysis, we provide insights that guide the development of more efficient automatic code optimization algorithms.
\end{abstract}


\shrink[1]
\section{Introduction}


Designing an autoscheduler—an algorithm that automatically determines the best sequence of code optimizations—remains challenging. Compiler developers must navigate many decisions when building these algorithms. For instance, should the autoscheduler allow transformations to be applied in any order, or should it enforce a fixed sequence? While exploring all orders of transformations can yield better performance due to the interactions between transformations, this increases the search space exponentially, making it infeasible for large programs. On the other hand, using a fixed order is more scalable, but may sacrifice performance. Quantifying these trade-offs is crucial.

Another key design question is whether certain transformations should be applied multiple times during the optimization process. Although applying transformations repeatedly can improve performance, it also expands the search space and increases the complexity of the process. Deciding which transformations benefit from multiple applications and which do not is critical for balancing optimization quality with computational efficiency.

Answering these questions is far from straightforward due to the inherent complexity of modern hardware and the interactions between code transformations. Traditional approaches rely heavily on expert intuition, but a data-driven methodology offers a promising alternative. By using large datasets of random programs and optimization sequences, we can statistically analyze the impact of different design choices, providing empirical insights that guide the development of more efficient automatic code optimization algorithms.

In this paper, we use a data-driven approach to answer multiple questions that arise when designing an autoscheduler. We do this by analyzing a large dataset of optimized programs. The goal of this analysis is to better understand the behavior of transformations and their relationships and to extract rules that could be integrated into an autoscheduler.



\shrink[0.5]
\section{Overview\label{overview}}

In this section, we provide a general overview of our contribution and methodology. In this work, we are interested in studying the following transformations: loop interchange, loop skewing, loop tiling, loop parallelization, and loop unrolling. 
We are interested in exploring the application of these transformations to optimize programs using Tiramisu\cite{baghdadi2019tiramisu, baghdadi2018tiramisu1,baghdadi2020tiramisuDNNDenseSparse}, a state-of-the-art polyhedral compiler. Our goal is to later use the insights that we gain in this study to improve the Tiramisu autoscheduler \textit{LOOPer} \cite{merouani2024looper}. While we focus our study on the previously mentioned transformations, our approach is general enough and can be applied on any other transformation as long as it is included in the dataset used in the study.
We perform our analysis on two datasets that we present in Sec. \ref{dataset}. These datasets are created by generating a large number of random programs. For each randomly generated program, a set of optimization sequences is sampled from the code optimization search space. 
In our study, we ask questions, and for each of them, we design a method to find an answer and present our findings. We present these questions in Sec. \ref{study1} and Sec. \ref{study2}.

 
For both sections, each of the questions that we asked is presented. Each of the subsections is organized as follows: it starts by providing the general context and why we are asking the question. It is followed by (a) the question, then (b) the specific methodology of the experiment that we did to answer the question. Next, we present (c) the results and finish by (d) a conclusion that discusses how the answer could help make an informed decision in designing a search algorithm.


\shrink[0.5]
\section{Datasets\label{dataset}}

Our study is based on two distinct datasets. The first, \textbf{Dataset A}, contains 29 million data points, and the second, \textbf{Dataset B}, consists of less data points. Both datasets cover a wide range of programs, exploring a vast space of transformation sequences\footnote{sequences of transformations}. While most of our analysis is based on Dataset A due to its larger size and diversity, we use Dataset B in Sections \ref{unrolling} and \ref{order}, where Dataset A is less suited to answering the specific questions raised. The key difference between the datasets is that Dataset B explores transformations in a random order and allows transformations to occur multiple times in a single schedule. If the dataset used in an experiment is not explicitly mentioned, it implies that \textbf{Dataset A} was used.

To generate these datasets, we created a large corpus of transformed Tiramisu programs, for which we measured and recorded speedups. The data generation process followed a two-step sampling approach: first, we sampled the program space by randomly generating synthetic Tiramisu programs, combining sequences of common computation patterns such as reductions and stencils. Then, we sampled the transformation space by collecting candidate transformations using a search technique (beam search as described by Baghdadi et al.\cite{tiramisu21}). During the exploration, each candidate schedule was applied, the transformed program was compiled and executed, and its speedup was measured and stored as a new data point in the dataset. We avoided random sampling of transformations as it would likely include unrealistic combinations, resulting in less useful examples. Our sampling method, which aligns with real-world use cases, ensures that the generated dataset reflects practical transformation scenarios encountered during code optimization.

The representativeness and generalization capacity of our synthetic dataset are further supported by prior research. This same dataset was previously used to train a deep learning model that successfully predicted speedups and guided beam search in unseen programs, demonstrating that it can generalize well to new data \cite{baghdadi2021deep}. This indicates that the dataset captures the essential characteristics of real-world optimization challenges, justifying its use in our study.

Moreover, using synthetic datasets instead of real-world datasets is justified by the need for control and diversity in data generation. Synthetic datasets allow us to systematically explore a wide range of program structures and transformations, enabling us to draw general-purpose conclusions from the statistical analysis. Real-world datasets, while valuable, are often limited in scope and fail to cover the broad spectrum of optimization scenarios necessary for a comprehensive analysis. Our controlled generation approach ensures that a representative sample of transformations is explored, leading to more robust and generalizable insights into code optimization.

The process of generating \textbf{Dataset A} took approximately eight months on a 15-node cluster, with each node featuring a dual-socket 12-core Intel Xeon E5-2695v2 CPU, 128 GB RAM, and an Infiniband interconnect. Generating \textbf{Dataset B} took a few weeks on a single-node machine equipped with an AMD EPYC Rome CPU with 128 cores and 512GB RAM.

\section{Statistical Exploration Of Code Transformations\label{study1}}

In this section, we will present and discuss the questions that we have explored along with our findings. The goal is to understand the behavior of the explored transformations and gain insights on how to efficiently explore the search space of transformations.

\shrink[1]
\subsection{Loop Parallelization}
Parallelizing the outermost loops in a loop nest is usually recommended. The autoscheduler usually tries to parallelize the outermost loop, and if it is not parallel, moves to try the next inner loop and so on. Exploring all the loop levels for parallelism increases the size of the search space, and therefore, minimizing the number of loops explored is desired.

\paragraph{Question}
At which loop depth should the exploration of parallelism stop? Should the compiler explore the parallelization of the outermost loop alone?

In other words, what is the loop depth k, such that exploring the k outermost loops for parallelism is enough, and exploring loops deeper than k would either lead to small speedups or to a slowdown.

\paragraph{Methodology} Given that different loop nests have different loop depths, it is hard to find a single value for k. Instead of looking for a single value for k, we propose to use \emph{relative loop levels}.
The relative loop level is the result of dividing the absolute loop level in a given loop nest on the depth of the loop nest. This gives a value between 0 and 1, where 0 represents the outermost loop and 1 represents the innermost loop, regardless of the depth of loop nests.

For each program in the dataset, the relative loop level that enables the best speedup has been collected to compute the mean of all speedups associated with that relative loop level.

\begin{figure}[htbp]
  \shrink[1.5]
  \centering
  \caption{Variation of Mean Speedup enabled by parallelization relatively to the relative loop level}
  \includegraphics[width=0.3\textwidth]{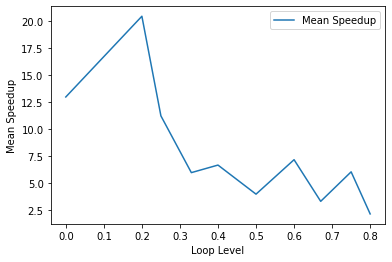}
  \label{fig:Figure 1}
  \shrink[1]
\end{figure}

\paragraph{Statistics and Results}
The results are in Figure~\ref{fig:Figure 1}.



As we can see from Figure~\ref{fig:Figure 1}, parallelizing the outermost loop alone (relative loop level = 0) without considering the inner loops is not the best option on average. Parallelizing the innermost loop is not ideal, either. Parallelizing the \emph{$30\%$} outermost loops yields speedups above $5\times$ on average. 

\paragraph{Answer}
It is safe to skip testing parallelism on the innermost loop level as it rarely yields speedups. To reduce the search space without sacrificing much speedup, the autoscheduler can focus on parallelizing the \emph{$30\%$} outermost loops since that yields the highest speedups on average.
This is in contrast to the common wisdom of exploring parallelization on the outermost loop alone.

\shrink[1]
\subsection{Loop Skewing}

Loop skewing is an affine transformation that changes the order of traversing iterations in the iteration domain. It is used to achieve two main goals: 1) eliminate loop-carried dependencies to enable parallelism; 2) maximize data locality. While skewing enables parallelism and data locality, it introduces an overhead due to the new shape of the loop. 

\paragraph{Question} First, does the use of skewing lead to speedups? (even though it introduces an overhead). Second, assuming we have a code that is parallel, would it be useful to explore skewing to also improve data locality?

\paragraph{Methodology}
While generating our dataset (Dataset A), and when exploring skewing, we used an algorithm similar to the Pluto\cite{bondhugula_practical_2008} algorithm to find the best skewing parameters. The algorithm returns two lists of parameters: the first maximizes parallelism, while the second maximizes data locality. Two analyses have been applied to the dataset to answer the previous questions.

The first analysis is to test the impact of skewing on the speedup without applying parallelization afterward, and with applying parallelization afterward. 

The second analysis is to check whether applying skewing before parallelization improves the speedup unlocked by parallelization or not.

\paragraph{Statistics and Results}
Applying parallelization after skewing enables a $3.662\times$ better average speedup than not applying parallelization after skewing. This shows that skewing, when combined with parallelization, leads to speedups. This goes against the thought that the overhead of skewing is higher than the benefit of parallelization and justifies the importance of exploring loop skewing in the search space.

The mean speedup of code where skewing was applied but parallelization was not is $1.169\times$. This shows that applying skewing with a goal other than parallelizing code (e.g., to maximize data locality) does not lead to significant speedups.



\paragraph{Answer}
Exploring skewing to enable parallelization is useful. And the overhead of skewing is small compared to the benefit of parallelization.
Exploring skewing to maximize data locality does not lead to high speedups and, therefore, can be dropped from the search space if there is a need.
A general rule to derive: if parallelization is not legal, explore skewing; otherwise, no need to explore skewing since it will not significantly improve data locality.

\shrink[0.5]
\subsection{Unrolling\label{unrolling}}

 While loop unrolling can provide performance benefits by reducing loop overhead and increasing instruction-level parallelism, choosing the right loop unrolling factor requires careful consideration of the tradeoff between these benefits and the side effects, which include an increase in code size and higher register pressure.

\paragraph{Question} Which unrolling factor should be used as the default unrolling factor for the CPUs that we consider?

\paragraph{Methodology}

In order to answer, we conducted the following experiment on both datasets,  \emph{Dataset A} and \emph{Dataset B}. We conducted this experiment on the two datasets since they were generated on different machines (different hardware) and therefore provide insights into how the default loop unrolling factor changes from one machine to another.

The dataset A contains three unrolling factors (4, 8, and 16). For both datasets, we categorize schedules based on these unrolling factors (4, 8, 16). Schedules that do not contain unrolling are ignored. We, then, compute the mean speedup associated with each unrolling factor.

\paragraph{Statistics and Results}

The following plots show the results of the experiment on both datasets.

\begin{figure}[htbp]
  \shrink[2]
  \centering
  \caption{Speedup variation based on unrolling factors - Dataset A}
  \includegraphics[width=0.25\textwidth]{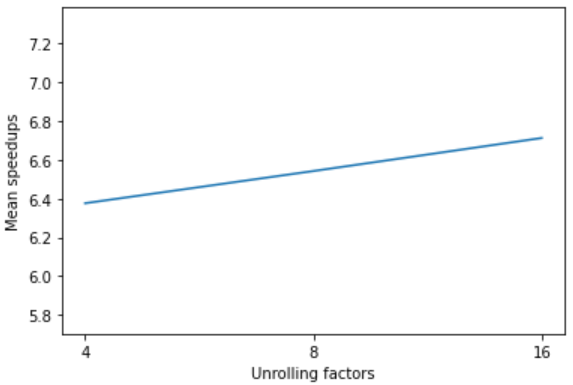}
  \label{fig:Figure 3}
  \shrink[3]
\end{figure}

\begin{figure}[htbp]
  \centering
  \caption{Speedup variation based on unrolling factors - Dataset B}
  \includegraphics[width=0.25\textwidth]{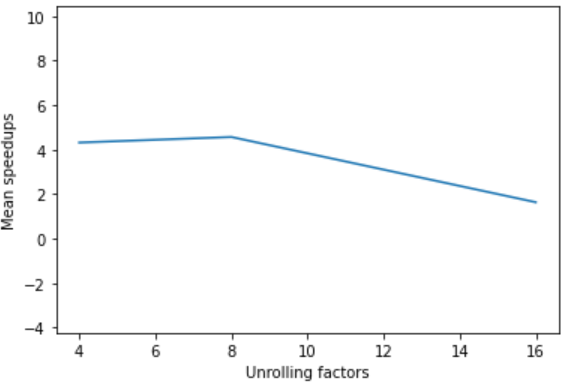}
  \label{fig:Figure 4}
  \shrink[1.5]
\end{figure}

In Fig. \ref{fig:Figure 3}, we can see that speedups increase as we increase the unrolling factor. The maximum speedup is reached with an unrolling factor of 16. Fig. \ref{fig:Figure 4} shows a slight increase in speedup as the unrolling factor increases, until the maximum speedup is reached (with an unrolling factor of 8). Increasing the unrolling factor beyond this point decreases speedups (which can be explained by a higher register pressure).

\paragraph{Answer} 
A default unrolling factor of 16 can be used for the machine used to generate Dataset A, and a default factor of 8 for the machine used for Dataset B. These default choices provide the best average speedup.


\shrink[1]
\section{Statistical Exploration of Schedule Properties\label{study2}}
\shrink[1]

We will analyze some of the properties of sequences of transformations. The purpose of these experiments is to help \emph{the autoscheduler} in reducing the search time without significantly impacting the quality of the final schedule.

\shrink[1]
\subsection{Schedule Length}

Allowing the autoscheduler to explore a long sequence of transformations allows for better chances to find the best speedups. While this is true, such exploration requires significantly more time. Also, if one were to build a machine learning cost model to guide the exploration, such a model needs to be trained on a larger dataset because the dataset needs to cover schedules with a long sequence of transformations as well as those with a short sequence of transformations.

\paragraph{Question} What should be the length of the sequence of transformations that an autoscheduler should explore?

\paragraph{Methodology}
To answer the previous question, we proposed two statistics:
For the first statistic, we counted the number of programs per schedule length in such a way that the best schedule in terms of speedup of a given program has that specific length.
In order to get more insights about the question, we also computed another statistic, which is both the mean and the maximum speedup per schedule length.

\paragraph{Statistics and Results} Figure 5 shows the results of the first statistics.

\begin{figure}[htbp]
  \shrink[1]
  \centering
  \caption{Number of functions per schedule length}
  \includegraphics[width=0.4\textwidth]{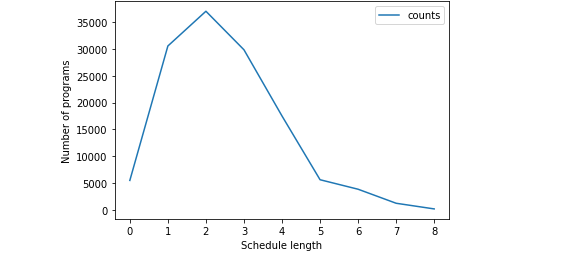}
  \label{fig:Figure 5}
  \shrink[2]
\end{figure}

 The second statistic's results are shown in Figure 6.

\begin{figure}[htbp]
  \shrink[2]
  \centering
  \caption{Mean and Maximal speedup by schedule length}
  \includegraphics[width=0.4\textwidth]{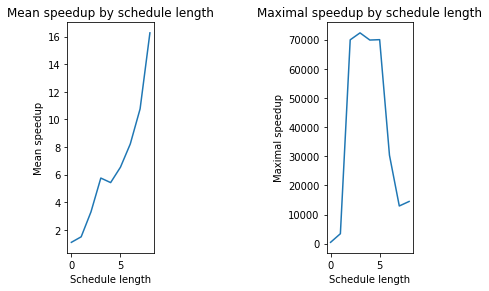}
  \label{fig:Figure 6}
  \shrink[1.5]
\end{figure}

From these two figures, we can see that schedules that have more than 5 transformations are rarely better than shorter schedules. We can also see that beyond 5 transformations, on average, the benefit of exploring other transformations diminishes compared to the cost of exploration.

\paragraph{Answer} A reasonable schedule length that allows a maximal speedup while minimizing the exploration time is a schedule that has a maximum of 5 transformations.

\shrink[1]
\subsection{Transformations Order\label{order}}

Allowing the autoscheduler to explore transformations in any order enables finding better schedules, but at the expense of slowing down the search (the size of the search space grows exponentially with the search depth, making the search space exploration impractical).
Exploring transformations in a fixed order has the advantage of keeping the search space small, but might miss certain optimization opportunities.

\paragraph{Question} Can we pick a data-driven order for transformations?

\paragraph{Methodology}

For the following experiment, we used the \emph{Dataset B} since it explores transformations in arbitrary order. \emph{Dataset A} explores transformations in a fixed order.

In order to find a better order to apply the transformations, we estimated for each transformation the probability of transitioning to all other transformations. We calculate the transition matrix (let us call it \emph{T}). For each transition \emph{T\textsubscript{ij}}, corresponds the mean speedup of all schedules containing the transition from \emph{transformation i} to \emph{transformation j}. Then, each element of the matrix is divided by the sum of its row in order to make valid probabilities which sum to one and represent the transition probabilities.

\begin{figure}[htbp]
  \shrink[1]
  \centering
  \caption{Transformations Transition Probabilities Heatmap}
  \includegraphics[width=0.4\textwidth]{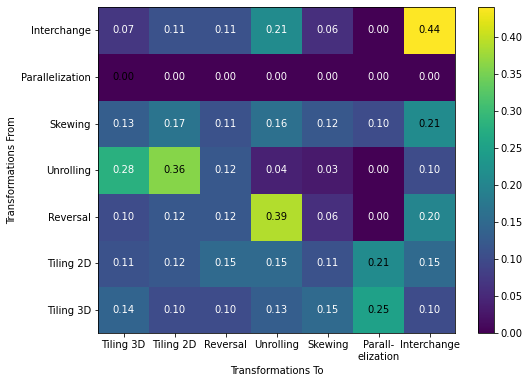}
  \label{fig:Figure 7}
    \shrink[1.5]
\end{figure}

\paragraph{Statistics and Results} To better visualize the matrix \emph{T}, we use a heatmap (presented in Fig.\ref{fig:Figure 7}). The horizontal axis represents the source transformation (in the transition), while the vertical axis represents the destination transformation. The transition with the highest probability for each transformation (each row) has been highlighted with a black font color.

In dataset B, parallelization is explored as the last transformation. Once the code is parallelized, the autoscheduler exits. The number of transformations explored before parallelization is unbound (the autoscheduler can explore any number of transformations in any order).

\paragraph{Answer}
From the transition matrix, we can see that applying interchange after interchange leads to the highest speedups on average (that is, it is useful to repeat the application of interchange multiple times before moving to another transformation).
Exploring interchange after skewing, and tiling after unrolling, and reversal after unrolling, and parallelization after tiling all lead to the best speedups on average. A possible data-driven fixed order based on this transition matrix would be: skewing, interchange*, reversal, unrolling, tiling, and parallelization. Although this would constitute the best default transformation order, it is not required to use all at once; some of the transformations can be skipped or repeated.

\shrink[0.5]
\section{Evaluation}
Our primary objective is to demonstrate the generalizability of our statistical-based rules across diverse, unseen functions within the Tiramisu autoscheduler \textit{LOOPer} \cite{merouani2024looper}. By integrating these rules into an existing beam search algorithm within \textit{LOOPer}, we aim to refine the decision-making process in automatic compilers, ensuring that they consistently identify high-quality transformation sequences. Through a series of experiments, we will evaluate the ability of these rules to maintain or enhance optimization performance across a wide range of functions. While our approach also seeks to reduce search time by pruning the search space, the emphasis is on showcasing the adaptability and broad applicability of our rules, ultimately illustrating their potential to streamline and improve the efficiency of the code optimization process without compromising results. 

\shrink[0.5]
\subsection{Search method: Beam search}
In the Tiramisu autoscheduler \textit{LOOPer} \cite{merouani2024looper}, beam search is employed to explore a search tree, where each branch represents a sequence of transformations. The algorithm iteratively expands the \textbf{K}-best branches at each level, with \textbf{K} (the beam size) defined as a search parameter. At the initial level, the autoscheduler generates a set of candidate transformations, evaluates each one using an evaluation function that quantifies the quality of each branch, and selects the top \textbf{K} candidates for further exploration. For each selected candidate, the same process of selection and expansion is repeated.

This exploration continues until no new candidates can be generated. Throughout the search, any transformations that violate data dependencies in the input program are detected and pruned using classical polyhedral dependence analysis and legality checking \cite{FEAUTRIER1988ARRAY, vasilache2006violated}. Additionally, at each level, the option of applying no transformation is included, which is beneficial when all new transformations would reduce performance. To prevent cycles, each generated candidate is enforced to be unique and is verified as unexplored in other branches of the search space.

  \shrink[0.5]
\subsection{Our statistical rules integration}
To evaluate the effect of our statistical-based rules, derived from our datasets, we will adjust the implementation of the Looper search strategy (which relies on beam search) and use the newly modified Looper to search for code optimizations for a standard benchmark.

\begin{enumerate}
    \item \textit{Parallelization:} Before generating a parallelization loop candidate, we compute its relative depth by dividing the loop’s depth by the total depth of its corresponding loop nest's abstract syntax tree. Parallelization is considered only if the loop falls within the outermost 30\% of loops.
    \item \textit{Skewing:} Rather than considering skewing candidates by default, skewing is only evaluated if parallelization is not legally viable for any of the shared computations within the loop nest. In cases where no parallelization is legal, skewing candidates are considered.
    \item \textit{Unrolling:} Rather than exploring multiple unrolling factors for innermost loops, we focus on the statistically favorable unrolling factor of \textbf{16}. As shown in Figure \ref{fig:Figure 3}, this factor provides the best performance as we observe a linear speedup relative to unrolling factors, as we adhere to the default transformation sequence used in \emph{Dataset A} for a first evaluation.
    \item \textit{Schedule Length:} Figures \ref{fig:Figure 5} and \ref{fig:Figure 6} indicate that overly lengthy schedules typically yield lower performance compared to more concise, well-designed schedules. Based on this observation, we limit the maximum schedule length to \textit{8 transformations} to extend the conclusion regarding optimal exploration length, previously capped at 5 transformations.
\end{enumerate}

Our rule (5.2) on transformation order could not be tested within the beam search of \textit{LOOPer}, as this autoscheduler does not accept different orders than the already implemented.

  \shrink[1]
\subsection{Benchmarks}
We utilized the PolyBench benchmark suite \cite{louis-noel_polybench_2010}, a widely recognized benchmark suite for assessing polyhedral compilers. PolyBench comprises 30 benchmarks derived from various computing domains, including linear algebra, stencils, and physics simulations. For each benchmark, we tested five problem sizes defined by PolyBench (MINI, SMALL, MEDIUM, LARGE, and EXTRALARGE) and used the default data types. To simplify the presentation of results, we calculated the geometric mean of the speedups achieved across all five sizes for each benchmark.
We performed the evaluation on a system running on a dual-socket 12-core Intel Xeon E5-2695v2 CPU equipped with 128GB of RAM, a similar experimental setup where \textit{LOOPer} has been evaluated with.

\subsection{Results}
\begin{figure}[htbp]
  \shrink[1]
 \centering
 \caption{Speedups of \textit{LOOPer} using the cost model with our statistical-based rules compared to the original \textit{LOOPer}. The speedups are aggregated by geometric mean over the five sizes of each benchmark. The benchmarks are sorted in descending order.}
 \includegraphics[width=0.4\textwidth]{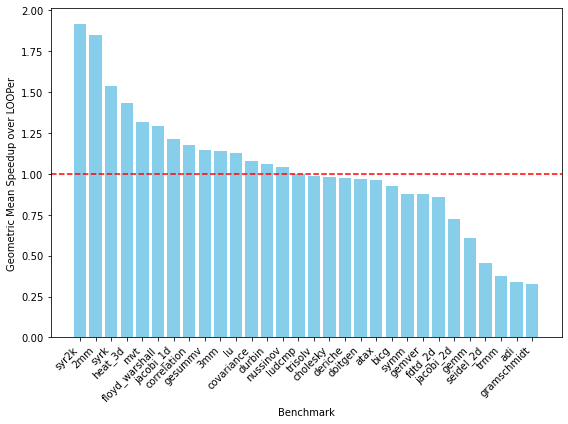}
 \label{geomean_speedup}
   \shrink[3]
\end{figure}

By integrating statistical-based rules into an autoscheduler named LOOPer, which optimizes code automatically, we achieved performance improvements for over 53\% of functions, while the next 30\% of the functions demonstrated results very close to baseline performance (see Figure \ref{geomean_speedup}). This was accomplished by adding three simple rules rooted in basic statistical measures, such as the mean. The overall arithmetic mean speedup reached 1.15x. Furthermore, we observed an improvement in search time for most functions, with an arithmetic mean of 1.24x. These findings highlight two key aspects: (1) the generalizability of our rules, as they proved effective even on functions not included in our dataset, thereby confirming their representativeness; and (2) the potential of data-driven code optimization. The integration of these three simple statistical rules based on the mean yielded notable performance gains, including a runtime improvement of nearly 25\%. We believe that further statistical analysis could deepen these results and lead to greater performance.

\section{Related Work}

Using ML for automatic code optimization has been widely explored~\cite{wang2018machine, ashouri2018survey,adams2019learning,baghdadi2021deep,merouani2020deep,chen2018learning,zheng2020ansor,mendis2019ithemal,polygym,merouani2024looper,hakimi2023hybrid,mezdour2023deep,bendib2024reinforcement,autophase2019,mlgo,compilergym,felix,looper_data_arxiv25,predictivepolyhedral}.
Tiramisu’s autoscheduler \cite{baghdadi2021deep,baghdadi2019tiramisu,merouani2024looper} performs tree-based search over polyhedral transformations, guided by a deep learning based cost model. In the same way, AutoTVM \cite{chen2018learning} employs a learned model to explore schedules using tree-search algorithms.

Automatic code optimization involves adapting compilers to optimize each program individually. 
Non data-driven automatic code optimization has been widely studied, with many approaches explored. One example of these approaches is the polyhedral model~\cite{feautrier_array_1988}. It is a mathematical model for representing code and code transformations and is one of the approaches used to automate compiler code optimization~\cite{Iri88,feautrier_array_1988,wolf1991loop,lefebvre_automatic_1998,Qui00,thies_unified_2001,Darte_contraction_2005,bondhugula_practical_2008, baghdadi2015PhD,baghdadi2019tiramisu,baghdadi2018tiramisu1,trifunovic_graphite_2010,polly,tobias_hexagonal_cgo13,Vasilache2018TensorCF,baghdadi2011speculation,merouani2020deep, pouchet.11.popl,baghdadi2020tiramisuDNNDenseSparse}.

\shrink[0.5]
\section{Conclusion}
\shrink[0.5]

Our work explores a data-driven approach to answer questions that designers of automatic code optimization algorithms usually face.
Among our findings
\begin{itemize}
    \item To reduce the search space without sacrificing speedup, the autoscheduler can focus on parallelizing the \emph{$30\%$} outermost loops since that yields the highest speedups on average. This is in contrast to the common wisdom of exploring parallelization on the outermost loop alone.
    \item The overhead of skewing is small compared to the benefit of parallelization that it enables. Exploring skewing to maximize data locality does not lead to high speedups and therefore can be dropped.
    In general, if parallelization is not legal, it is useful to explore skewing, otherwise, there is no need to explore skewing.
    \item A reasonable schedule length  that allows a maximal speedup while minimizing the exploration time is a schedule that has a maximum of 5 transformations;
    \item We quantified the transition probabilities between pairs of transformations which allows a data-driven selection of a fixed order for exploring transformations.
\end{itemize}

The data-driven rules that we derived in this paper can be easily integrated in an autoscheduler. We tested our rules with \textit{LOOPer} on PolyBench Benchmark Suite, and we could observe an improvement in the performance for most functions while reducing search time of the autoscheduler. This demonstrates not only the representativeness of our datasets but also the effectiveness of our rules.

\shrink[0.5]
\section{Acknowledgment}
\shrink[0.5]

This research has been partly supported by the Center for Artificial Intelligence and Robotics (CAIR) at New York University Abu Dhabi, funded by Tamkeen under the NYUAD Research Institute Award CG010. The research was carried out on the High-Performance Computing resources at New York University Abu Dhabi.

\bibliographystyle{unsrt}
\bibliography{sample-base}

\end{document}